\documentclass[aps,prb,showpacs,twocolumn,superscriptaddress]{revtex4-1}
\usepackage{graphicx,subfigure}
\usepackage{color}
\begin{document}


\title{The R\'enyi Entropy and the Multifractal Spectrum of Systems Near the Localization Transition }

\author{Xiao Chen}
\affiliation{Department of Physics, University of Illinois at Urbana-Champaign, 
Illinois 61801-3080, USA}

\author{Benjamin Hsu}
\affiliation{Department of Physics, Princeton University, Princeton, NJ 08544, USA}

\author{Taylor L. Hughes}
\affiliation{Department of Physics, University of Illinois at Urbana-Champaign, 
Illinois 61801-3080, USA}

\author{Eduardo Fradkin}
\affiliation{Department of Physics, University of Illinois at Urbana-Champaign, 
Illinois 61801-3080, USA}

\date{\today}

\begin{abstract}
We show that the R\'enyi entropies of single particle, critical wave functions for disordered systems contain information about the multifractal spectrum. It is shown for moments of the R\'enyi entropy, $S_{n}$, where $\vert n\vert<1$, it is possible to extract universal information about the multifractility of such systems. This is shown through a generic calculation and then illustrated through two example models. We find good agreement between our analytic formula and numerical simulations of the two test models. Our formalism is easily extendable to generic non-interacting fermion models.  It is also suggested that recent experimental advances in measuring the multifractal spectrum might allow some moments of the R\'enyi entropy to be measured.
\end{abstract}


\maketitle

\section{Introduction}

There has been a recent influx of ideas and tools from quantum information theory that are being used to characterize many-body condensed matter systems. Part of this interest stems from the fact that some subtle properties of the underlying phase of matter, otherwise difficult to calculate, can be found quite naturally by partitioning the density matrix of a ground-state wave function and calculating measures of entanglement. For example, in one dimensional critical systems, it was shown that the central charge can be determined via the scaling of the entanglement entropy which takes the form, $S \sim c/3 \log L$ where $L$ is the size of the partitioned region and $c$ is the central charge characterizing the critical point\cite{Calabrese2004}. In higher dimensional critical systems general results remain unknown, though interesting behavior has been found in a few isolated critical systems.\cite{Fradkin2006,oshikawa,Marie2009,Hsu2009, Hsu2010, Metlitski}

Entanglement finds a natural place in the study of disordered systems where the concepts of entanglement and de-localization go hand in hand. For example, the entanglement entropy at  infinite-randomness fixed points displays a similar $\log L$ scaling behavior in which, by analogy, it was proposed that these models may have very unconventional values for the central charge\cite{Rafeal2009}. It was also conjectured that while the entanglement entropy has a similar behavior as the translationally invariant case, the R\'enyi entropies are very different, and in fact constant. Recent studies of the so-called \emph{single-particle} entanglement entropy in disordered systems, in contrast to the many-body entanglement entropy described above, have found an interesting scaling behavior, $S \sim \alpha_{1} \log L$, where $L$ is the system size and $\alpha_{1}$ is related to the multifractal spectrum through, $\alpha_1 = \partial_{n} \tau(n) \vert_{n=1}$ where $\tau (n)$ are the scaling exponents of the inverse participation ratios (IPRs) (defined below)\cite{Charkravarty2010,Charkravarty2008}. In this work we extend this result and study the R\'enyi entropy for non-interacting disordered systems. Whereas references \onlinecite{Charkravarty2010,Charkravarty2008} only examined the $n=1$ moment of the R\'enyi entropy $S_{n}$, we show that the other moments yield more information about the multifractal spectrum. Indeed more than the derivative of the multifractal spectrum at $n=1$ can be found by looking at other moments. Specifically, for small $n$  the scaling behavior of the R\'enyi entropy and the IPRs are identical. Since the multifractal spectrum for the critical wavefunction is universal at Anderson transition, the R\'enyi entropy for small $n$ is also universal \cite{Kadanoff1987, Chamon1997}.  Thus we show that as in the case of critical one dimensional systems, universal information can indeed be found through entanglement.

The multifractal spectrum (MFS)\cite{Halsey-1986,Kadanoff1987}, like the central charge for critical 1+1 dimensional systems, plays an important role in characterizing non-interacting disordered systems\cite{Wegner1980,Duplantier1991}. More concretely, the MFS is the set of scaling exponents of the IPR, defined as $P_{n} = \sum_{i} \vert \psi_{i}(x) \vert^{2n}$ which is indexed by a real number $n.$ It has been observed that in many critical disordered systems the IPRs have a spectrum of scaling exponents defined by $P_n\sim L^{-\tau(n)}$ characterized by the spatial dimension $d$ and function $\tau(n)$ which is the MFS. Generically, the MFS can be written as $\tau(n) = D_{n}(n-1)$ where $D_{n}$ is a function of $n$ that depends on the system. For a conventional band insulator $D_{n}= 0$ while in a metal $D_{n}=d$. At a critical point, the $D_{n}$ can have a non-trivial dependence on $n$\cite{Mirlin2007}. The MFS and its symmetry properties play a central role in restricting the possible types of critical field theories one can construct\cite{Chamon1996,Gruzberg2011,Duplantier1991,Mudry1996}. An understanding of the multifractal nature of critical wave functions is hence important for an understanding of the localization transition. Additionally, advances in imaging have also made it possible to begin to observe the multifractal spectrum in a number of disordered systems at the Anderson transition\cite{Chabe2008,Faez2009,Hu2008,Morgenstern2003}. Our interest, however, in the connection between the MFS and the R\'enyi entropies,  is chiefly theoretical. While for the sake of argument we consider models where the multifractal spectra are known, in general this set of exponents is difficult to compute. We demonstrate that a much simpler calculation gives an accurate approximation to an interesting region of the MFS.  By an explicit calculation and two numerical examples, we show that the information content of critical wavefunctions of disordered free-fermion models contains imprinted signatures of their MFS.

\section{R\'enyi Entropies}

The entanglement measures we focus on are the standard von-Neumann entanglement entropy and a closely related quantity, the R\'enyi entropy. Given a wave function $\vert \Psi \rangle$, it is possible to construct the density matrix as, $\hat{\rho}= \vert \Psi \rangle \langle \Psi \vert$. The conventional entanglement entropy is calculated by partitioning the system into two spatial subregions $A$ and $B$, where only one subregion is observed, say region $A$. One then traces out all of the possible configurations of region $B$ to yield the reduced density matrix $\hat{\rho_{A}} = \textrm{Tr}_{B}\hat{\rho}.$ The question is then asked how much information does $A$ know about region $B$ and an obvious candidate for such a measure is the von-Neumann entropy of region $A$, $S=-\textrm{Tr } \hat{\rho}_{A} \log \hat{\rho_{A}}$ \emph{i.e.} the entanglement entropy. A closely related quantity is the set of R\'enyi entropies, 
\begin{equation}
	S_{n} = \frac{1}{1-n} \log \textrm{Tr }\hat{\rho}_{A}^{n}.
\end{equation}\noindent parameterized by a real number $n$ such that as $n\to 1$  the R\'enyi entropy reduces to the von-Neumann entropy.

We are focusing on localization in terms of single particle physics and thus we will use a notion of single-particle entanglement for a choice of a single-particle eigenstate. One can define entanglement using a site occupation number basis in the second quantized Fock space\cite{Zanardi2002}. From now on, we work with a single-particle eigenstate $\psi(x)$ which can be written in the single particle occupation basis as done by previous authors\cite{Charkravarty2010, Charkravarty2008}. Explicitly, the wave function can be written as 
\begin{equation}
	\vert \psi \rangle = \sum_{r\in A \cup B} \psi(r) \vert 1 \rangle_{r} \otimes_{r\neq r'} \vert 0 \rangle_{r'},
\end{equation}
where $\psi(r)$ is the normalized single particle wavefunction and $\vert n\rangle_{r}$ is the state with $n$ particles on site $r$. The wavefunction can then be written as,
\begin{equation}
 	\vert \psi \rangle = \sum_{ij} M_{ij} \vert 1 \rangle_{r_{i}\in A} \otimes \vert 1 \rangle_{r_{j}\in B},
\end{equation}
where $\vert 1\rangle_{r_{i}\in A}$ are states where the single particle is located in region $A$ at site $r_{i}$. The matrix $M_{ij}$ generically has the form,
\begin{equation}
	M_{ij} = \left( \begin{array}{cccc} 0 & b_{1} & b_{2} & \dots \\
							a_{1} & 0 & 0 &\dots \\
							a_{2} &0 &0 &\dots \\
							. & 0 & 0 & \dots \\
							. & 0 & 0&\dots  \end{array} \right).
\end{equation}
$a_{i},b_{i}$ denoting the amplitude of the wavefunction at site $i$ in region $A,B$ respectively. The Schmidt decomposition of such a wavefunction is easily carried out. Using singular value decomposition, we can define a new basis $u^{\dagger}_{i j} \vert j \rangle_{A}$ and $v_{ij}\vert j \rangle_{B}$ where the matrices $u,v$ are the unitary matrix of singular vectors for the matrix $M$. That is, they satisfy the relationship
\begin{equation}
	M_{ij} = u^{\dagger}_{ik} \lambda_{kk} v_{kj},
\end{equation}
where $\lambda_{kk}$ is the diagonal matrix containing the singular values of the matrix $M_{ij}$. There are only two non-zero elements in the single particle problem: $\lambda_{1} = \sqrt{ \sum_{i}b_{i}^{2} }$ and $\lambda_{2}=\sqrt{\sum_{i}a_{i}^{2}}$. The reduced density matrix $\hat{\rho}_{A}$ is then easily written since the density matrix is now diagonal. One simply finds that
\begin{equation}
	\textrm{Tr } \hat{\rho}_{A}^{n} =\left[\sum_{i\in A}a_{i}^2\right]^n+\left[\sum_{i\in B}b_{i}^2\right]^n=p_{A}^n+p_{B}^n
\end{equation}\noindent where $p_{A/B}$ is the probability that a particle in the single-body wavefunction $\vert\psi\rangle$ lies in region A or B respectively. Hence, we may write the R\'enyi entropy as
\begin{equation}
	S_{n} = \frac{1}{1-n} \log \left( p_{B}^{n} + p_{A}^{n} \right),
\end{equation}
which agrees with the expression for the entanglement entropy obtained by Ref. \onlinecite{Charkravarty2008} as $n\to 1.$

Since we have restricted our attention to the one particle sector of the density matrix and not the full many-body density matrix, the entanglement entropy should not grow large with the subsystem size as has been noted previously\cite{Charkravarty2008,Zanardi2002}. The expectation is that the same is true here. All the interesting behavior is in the scaling with the total system size. Indeed, by looking at the entanglement of a single site, we can extract the dominant scaling behavior with the system size, $L$. Looking at the case where region $A$ is a single site $x_{i}$, we simply have $p_A(x_i)=x_{i}^d$ where $d$ is the dimension. Clearly, any site $x_{i}$ is not especially privileged and hence, one should look at the average over all possible sites, chosen with a uniform probability distribution. Denote $[\cdot]_{x_{i} }$ as this site average with respect to the uniform probability distribution. That is, in $d$ dimensions each site $x_{i}$ is chosen with equal probability $(a/L)^{d}$ where $a$ is the lattice constant. If we look at \emph{the site averaged} $\textrm{Tr }\hat{\rho}_{A}^{n}$ we obtain,
\begin{eqnarray} \label{eqn:Tr_rho_n}
	&[ \textrm{Tr }\hat{\rho}_{A}^{n} ]_{x_{i} }& =   \left( \frac{a}{L} \right)^{d} \ \sum_{x_{i}} \left( (1-p_{A}(x_i))^{n} +p_{A}^{n}(x_i) \right) \nonumber \\
	&=& \left( \frac{a}{L} \right)^{d} \left [ \sum_{m=0}^{\infty} \left( \begin{array}{c}  n \\ m \end{array} \right) (-1)^{m} \sum_{x_{i} }p_{A}^{m} + \sum_{x_{i} } p_{A}^{n} \right] \nonumber \\
          &=& \left( \frac{a}{L} \right)^{d} \left[ \sum_{m=0}^{\infty} \left( \begin{array}{c} n\\m \end{array} \right) (-1)^{m} P_m+P_n\right] \nonumber\\
          &\sim& \left( \frac{a}{L} \right)^{d} \left[ \sum_{m=0}^{\infty} \left( \begin{array}{c} n\\m \end{array} \right) (-1)^{m} \left(\frac{a}{L}\right)^{\tau(m)}+\left(\frac{a}{L}\right)^{\tau(n)}\right] \nonumber\\
\end{eqnarray}
The quantity $\sum_{x_{i} } p_{A}^{n}(x_i)$ can be identified with the inverse participation ratio, $P_{n}$ which we recall scales as  $P_{n} \sim \left( a \slash L \right)^{\tau(n)}$, where the $\tau(n)$ are the multifractal spectrum.  One finds that for $ \vert n\vert  <  1 $ the multifractal spectrum is strictly less than zero and the site-averaged $\textrm{Tr}\rho_{A}^{n}$ can be approximated as,
\begin{equation}\label{eqn:trp}
	[ \textrm{Tr }\hat{\rho}_{A}^{n} ]_{x_{i}} \simeq 1 - n(a/L)^{d} + \left(a/L\right)^{\tau(n)+d}.
\end{equation}
By extension, the site averaged R\'enyi entropy can be approximated as,
\begin{equation}\label{eqn:result}
	(1-n) S_{n} \sim  -n\left(\frac{a}{L}\right)^{d} + \left( \frac{a}{L}\right)^{\tau(n)+d}, \quad \vert n\vert < 1
\end{equation}
which is the main result of this work and clearly shows the relationship between the R\'enyi entropies and the multifractal spectrum. Grassberger\cite{Grassberger-1983} found a related connection between  R\'enyi entropies and the multifractal spectrum of the probability measures of classical strange attractors.

As $n\rightarrow 1$, as expected, our result reproduces the result of Jia et. al \cite{Charkravarty2008} for the von Neumann entropy, up to a subleading correction that vanishes as $L\rightarrow \infty$. Indeed, there is always a region where $\tau(n)$ is less than zero where our approximation is valid since the multifractal spectrum is a non-decreasing, convex function with $\tau(0)=-d$, the spatial dimension, and $\tau(1)=0$. For moments $\vert n \vert >1 $, such a simplification can not be achieved and according to Eq.(8), we need to sum up all the terms to get the R\'enyi entropies. The connection between the moments of the R\'enyi entropies and the multifractal spectra for these large moments is less clear. We will now consider two well-studied models where the multi-fractal spectrum is known and compare our approximate form with the known results.

\section{Random Flux Model}
We first consider a system of Dirac fermions confined to 2d in the presence of a quenched random magnetic field normal to the plane with vanishing magnetic flux on average. Such a system can be thought of as the continuum limit of various tight binding Hamiltonians like the Chalker-Coddington network model\cite{Chalker-1988}. General considerations have predicted that the random magnetic field Dirac model has an exact zero-mode critical wave function which is multi-fractal.
As has been noted by numerous authors, the most useful property of this model is that the wavefunction can be calculated \emph{exactly} for any realization of the random magnetic field $B(x) = \nabla^{2}\Phi(x)$ and is given by $\psi(x) = Z^{-1} \exp(-\Phi(x))$ where $Z$ is the normalization factor. Here the random magnetic field $B(x)$ is assumed to be Gaussian distributed with variance (disorder strength) $g$, {\it i.e.} $P[\Phi] \propto \exp[-\frac{1}{2g} \int d^{2}x [\nabla\Phi]^{2} ]$\cite{Chamon1996,Chamon1997,Ryu2001}. To construct the reduced density matrix we simply note
\begin{equation}
	p_{A}(x_{i} ) = e^{-2\Phi(x_{i}) }\bigg\slash Z.
\end{equation}

\begin{figure}
	\includegraphics[width=.5\textwidth]{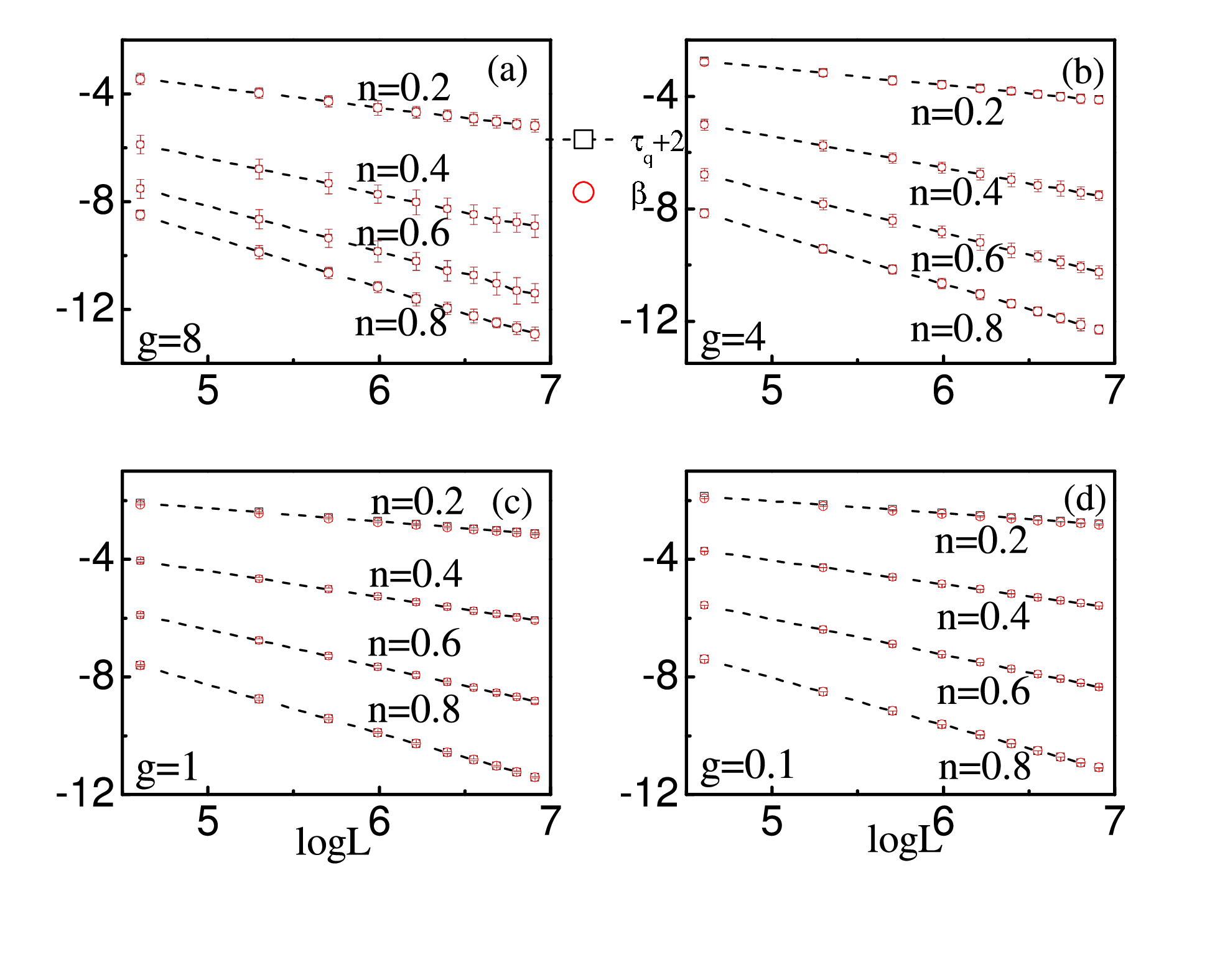}
	\caption{Comparison between the scaling behavior of $(1-n)S_{n} + n L^{-2} \sim L^{\beta}$ and the multifractal spectrum $\tau_{q_{c}}(n)+2$ for the random-flux Dirac model for system sizes L=[100,1000]. In figures (a), the system is in the strong disorder regime while in figures (b),(c),(d) the system is in the weak disorder regime. In both, we find good agreement between the scaling of the R\'enyi entropy and multifractal spectrum for exponents $\vert n\vert <1$.\label{fig2:} } 
\end{figure}

Depending on the value of $g$, its been shown that the multifractal spectrum for a Dirac fermion in a random magnetic field has two regimes. In the weak disorder regime, defined by $\sqrt{2\pi/g}\equiv q_{c}>1$, the multifractal spectrum displays a discontinuous behavior as a function of $n$ that was found in Ref. [\onlinecite{Chamon1996}]: 
\begin{equation}
	\tau_{q_{c}}(n ) = \Bigg\{ \begin{array}{ll} 2(1- \textrm{sign}(n)/q_{c} )^{2} n, &  q_{c} < \vert n \vert \\
								      2(1- n/q_{c}^{2} ) (n-1), & \vert n\vert \leq q_{c} \end{array} .
\end{equation}
In the strong disorder regime, where $q_{c}<1$, the similarity between the R\'enyi entropy and the multifractal spectrum can also be seen for values of the exponents given by
\begin{equation}
	\tau_{q_{c}}(n) =\Bigg\{ \begin{array}{ll} \frac{4}{q_c}(n-|n| ) &  q_{c} < \vert n \vert  \\
								-2(1-n/q_{c} )^{2}, & \vert n \vert \leq q_{c} \end{array} .
 \end{equation}
In both the weak and strong disorder regimes Eq. \ref{eqn:result} is applicable as long as $n<1.$
 In Figure \ref{fig2:}, we plot a few values of $n$ and compare the scaling behavior of the R\'enyi entropy and the multifractal spectrum in the strong disorder and weak disorder regimes for system sizes $L=[100,1000]$. Good agreement between the two quantities is found.

\begin{figure}[th]
\centering
\includegraphics[scale=.5]{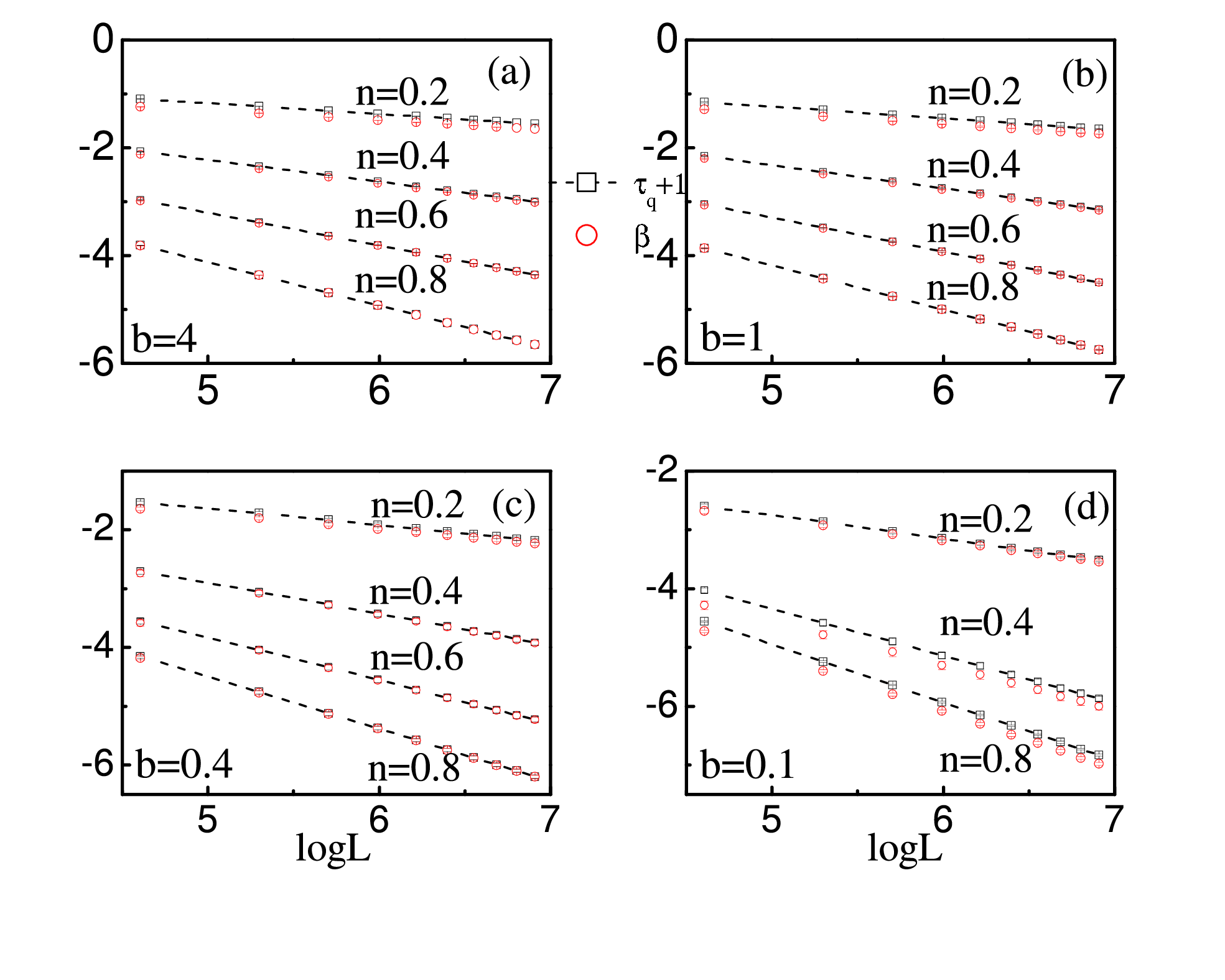}
\caption{The multifractal spectrum and the scaling of the R\'enyi entropy moments as a function of $n$ in the PRBM. The scaling behavior is shown for $n=0.2, 0.4, 0.6$ and $n=0.8$ for system sizes L=[100, 1000]. In figures (a) and (b), the systems is in the weak multifractal regime with $b=4$ and $b=1$  respectively. In figures (c) and (d), where $b=0.4$ and $b=0.1$ respectively, we can see that the agreement is good in the strong multifractal regime. In figure (d), only $n=0.2, 0.4$ and $0.8$ are shown for the sake of clarity. }
\label{fig1:}
\end{figure}

\section{Periodic-Random Banded Model}

To see that Eq. \ref{eqn:result} is quite general for any single particle wave function, we explicitly check our result in a much simpler model:  the periodic random banded model (PRBM). 
The model is defined as the ensemble of random Hermitian $L\times L$ matrices where the entires $H_{ij}$ are independently distributed Gaussian variables with mean zero and a variance that falls off as
\begin{equation}
	\langle \vert H_{ij} \vert^{2}\rangle = \frac{1}{1+(\vert i -j\vert/b)^{2\alpha} }.
\end{equation}
At $\alpha=1$ the model undergoes an Anderson transition from the localized $\alpha>1$ to delocalized $\alpha<1$ phases, for all values of $b$. At this point, the model shows  key features of the Anderson critical point, namely eigenfunctions possessing multifractal behavior. Again there is a single extended wave function on which we will  focus. We expect and confirm that  the previous analysis holds for moments less than $n<1.$ However, unlike the previous case, an analytic expression for the wave function is not known so we match our approximation to the numerical calculation of IPR scaling.  To eliminate the effect of boundaries, we will look at the periodic version of this model where
\begin{equation}
	\langle \vert H_{ij} \vert^{2}\rangle  = \left( 1+ \frac{1}{b^{2} } \frac{\sin^{2}(\pi r /N ) }{(\pi/N)^{2} } \right)^{-1}.
\end{equation}
This can be interpreted as describing a 1D model with long range hopping that falls off as $1/r^{\alpha}$ \cite{Mirlin2007}. From our previous results, it is expected that the combination $(1-n)S_n+n\frac{a}{L}$ should scale with the system size to some power, $ (\frac{a}{L})^{\beta}$, where $\beta$ is predicted to be $\tau(n)+1$ (note the change from $\tau(n)+2$ to $\tau(n)+1$ when compared with the random flux case due to the change in spatial dimension).  Systems of size L=[100, 1000]  were considered. As shown in Fig. \ref{fig1:}, $\beta$ matches very well with $\tau(n)+1$. In both the strong multifractal regime and weak multifractal regimes there is a range of values for which the scaling of the R\'enyi entropy and the inverse participation ratio are nearly identical.

\section{Conclusions}

In conclusion, we showed that looking at the extended wave functions in the single particle occupation basis yields a general relationship between the R\'enyi entropies and multifractal spectrum.  Our results can be applied to a wide-range of models, two of which were explicitly shown in this work. We hope that the R\'enyi entropy formula offers a simpler and more efficient way to calculate universal pieces of the multifractal spectrum.  The connection also opens the door to measuring R\'enyi entropies experimentally through the measurement of the multifractal spectrum via imaging. Finally, it would be interesting to explore to what extent the R\'enyi entropy and the multifractal spectrum are related for many-body ground state wave functions.

{\it Acknowledgments }-- This work was supported in part by the National Science Foundation through the grants DMR 0758462 (EF,TLH) and  DMR-1064319 (EF) at the University of Illinois. BH was supported by NSF grant PHY-1005429.

\end{document}